\documentclass[seceq,supplement]{ptptex}
\usepackage{amsmath}

\title{Response Functions for a Granular Fluid}
\author{James \textsc{Dufty}}
\inst{Department of Physics, University of Florida, Gainesville, FL
32611 USA} \abst{The response of an isolated granular fluid to small
perturbations of the hydrodynamic fields is considered. The
corresponding linear response functions are identified in terms of a
formal solution to the Liouville equation including the effects of
the cooling reference state. These functions are evaluated exactly
in the asymptotic long wavelength limit and shown to represent
hydrodynamic modes. More generally, the linear granular
Navier-Stokes equations for the response functions and related
Langevin equations are obtained from an extension of Mori's
identity. The resulting Green-Kubo expressions for transport
coefficients are compared and contrasted with those for a molecular
fluid. Next  the response functions are described in terms of an
effective dynamics in the single particle phase space. A closed
linear kinetic equation is obtained formally in terms of a linear
two particle functional. This closure is evaluated for two examples:
a short time Markovian approximation, and a low density expansion on
length and time scales of the mean free time and mean free path. The
former is a generalization of the revised Enskog kinetic theory to
include velocity correlations. The latter is an extension of the
Boltzmann equation to include the effects of recollisions (rings)
among the particles.}

\begin{document}

\maketitle

\section{Introduction}

It is generally accepted that many states of driven (activated) granular
matter admit a macroscopic description analogous to that of molecular
fluids. Such states will be referred to as "fluidized". As with normal
fluids, that continuum description can be simple or complex depending on the
particular fluid considered and its state (e.g., large or small spatial
gradients; near or far from boundaries; laminar or turbulent; homogeneous or
heterogeneous) \cite{General,Poschel}. In all cases, the basis for the
macroscopic continuum description is the exact local balance equations for
the relevant hydrodynamic fields, and suitable approximations for the fluxes
in these equations (constitutive equations). These approximations may be
based on extrapolations from experimental observations, constrained by
symmetries and other known theoretical properties. The simplest case is that
of a one component system for which the macroscopic fields are the density,
temperature, and flow velocity. If in addition, the dimensionless spatial
gradients and time derivatives are small then a Navier-Stokes hydrodynamics
is expected to apply. However, the form of these equations and the
dependence of their coefficients on grain properties and fluid state
conditions are not known \emph{a priori}. For granular gases, the
hydrodynamic equations and explicit expressions for their coefficients can
be derived from the granular Boltzmann equation. The origins of the
Boltzmann equation \cite{Dufty00} and a hydrodynamic description \cite%
{Origins} can be made quite clear in this case.

At higher densities, still within the constraints of small gradients, the
methods of linear response can be applied to obtain the linear Navier-Stokes
equations and formally exact expressions for the transport coefficients \cite%
{Aparna}; the generalization to obtain non-linear hydrodynamics in this way
has also been described \cite{DuftyReviews}. The application of linear
response parallels closely its application for the same purpose to molecular
fluids \cite{Hansen,McL}, although significant differences appear due to the
inelasticity of collisions for granular fluids. The next two sections give a
brief review of the central ideas to allow the reader to see these
similarities and differences. For example, an new concept of global
invariants is observed with a direct connection to long wavelength
hydrodynamics at the microscopic level. A new formal approach to writing
equations for the response functions is given in Section 4 from an extension
of the Mori projection operator method. Navier-Stokes hydrodynamics and
expressions for the transport coefficients are extracted from the long
wavelength form of these equations. In the second half of this presentation
the practical evaluation of the response functions is considered in the
context of kinetic theory. As observed elsewhere \cite{Aparna2}, a
non-trivial generalization (e.g., retaining velocity correlations) of the
phenomenological Enskog kinetic theory follows directly from a short-time
(Markovian) approximation. Next, the results of a systematic low density
expansion are described for leading order corrections to the Boltzmann
limit. The relevant length scale chosen in this expansion is the mean free
path, leading to the inclusion of "ring collisions" at this order. Some
outstanding problems and opportunities for the future are mentioned in the
summary section.

The results described here are a personal perspective developed primarily by
the author and his collaborators. The references reflect this bias, and
apologies are offered to those whose excellent contributions to related
topics are only recognized in the books and reviews cited here.

\section{Liouville equation and hydrodynamic modes}

\label{sec2}Attention will be restricted to the simplest realistic model of
a granular fluid: a system of $N$ smooth, inelastic, hard spheres of mass $m$%
, diameter $\sigma $, and positive normal restitution coefficient $\alpha
\leq 1$. The state of the system is specified by distribution function $\rho
\left( \Gamma \right) $ for the $N$ particle phase space $\Gamma =\left\{
\mathbf{q}_{1},..\mathbf{q}_{N},\mathbf{v}_{1},..\mathbf{v}_{N}\right\} $.
The evolution from a given initial state is given by the Liouville equation%
\begin{equation}
\partial _{t}\rho \left( \Gamma ,t\right) +\overline{L}\rho \left( \Gamma
,t\right) =0.  \label{2.1}
\end{equation}%
The operator $\overline{L}$ generates the hard sphere Liouville dynamics for
the $N$ particles: linear trajectories until a given pair is in contact,
followed by an inelastic change in that pair's velocities, and that process
continued. On each collision there is a loss of energy so the total energy
for an isolated system decreases in time \cite{Brey97}. It is usual to
define a temperature in terms of that energy by $E(t)\equiv 3NT(t)/2$. This
is simply a definition and has no implications for an underlying
thermodynamics. The decrease of $T(t)$ is represented by a cooling rate $%
\zeta \equiv -\partial _{t}\ln T$.

For an isolated system a special homogenous solution is sought in which all
of the time dependence occurs through the energy, or temperature
\begin{equation}
\rho _{h}\left( \Gamma ;n,T_{h}\left( t\right) ,\mathbf{U}\right) =\left(
\ell v_{0}(t)\right) ^{-3N}\rho _{h}^{\ast }\left( \left\{ \frac{\mathbf{q}%
_{\alpha }-\mathbf{q}_{\beta }}{\ell },\frac{\mathbf{v}_{\alpha }-\mathbf{U}%
_{\alpha }}{v_{h}(t)}\right\} ,n\ell ^{3}\right) ,  \label{2.4}
\end{equation}%
where $\ell $ is a characteristic length scale (e.g., mean free path) and $%
v_{0}(t)=\sqrt{2T_{h}(t)/m}$ is the thermal velocity. This is known as the
homogeneous cooling state (HCS), and is parameterized by the average density
$n$, the temperature $T_{h}(t)$, and the average system velocity $\mathbf{U}%
_{\alpha }$, all of which are uniform. The $\Gamma $ dependence of the
dimensionless distribution $\rho _{h}^{\ast }$ follows from the Liouville
equation%
\begin{equation}
\overline{\mathcal{L}}\rho _{h}^{\ast }=0,\hspace{0.25in}\overline{\mathcal{L%
}}X\equiv \overline{L}X+\frac{1}{2}\zeta _{h}\sum_{\beta =1}^{N}\nabla _{%
\mathbf{V}_{\beta }}\cdot \left( \mathbf{V}_{\beta }X\right) .  \label{2.6}
\end{equation}%
with $\mathbf{V}_{\alpha }=\mathbf{v}_{\alpha }-\mathbf{U}_{\alpha }$. In
the definition of $\overline{\mathcal{L}}$ the cooling rate $\zeta _{h}$
must be determined self-consistently, i.e.%
\begin{equation}
\int d\Gamma ^{\ast }K\overline{\mathcal{L}}\rho _{h}^{\ast }=0\Rightarrow
\zeta _{h}=\frac{2}{3n_{h}T_{h}}\int d\Gamma ^{\ast }\overset{\cdot }{K}\rho
_{h}^{\ast },  \label{2.6a}
\end{equation}%
where $K$ is the total kinetic energy and $\overset{\cdot }{K}$ is its time
derivative.

For more general states a similar dimensionless distribution can be defined%
\begin{equation}
\rho \left( \Gamma ,t\right) \equiv \left( \ell v_{0}(t)\right) ^{-3N}\rho
^{\ast }\left( \left\{ \frac{\mathbf{q}_{\alpha }}{\ell },\frac{\mathbf{v}%
_{\alpha }-\mathbf{U}_{\alpha }}{v_{h}(t)}\right\} ,n\ell ^{3},s\right) .
\label{2.7}
\end{equation}%
A corresponding dimensionless time $s$, $\overline{\mathcal{L}}$ and $\zeta
_{h}$ are given by%
\begin{equation}
s=\int_{0}^{t}dt^{\prime }\frac{v_{h}(t^{\prime })}{\ell },\hspace{0.25in}%
\overline{\mathcal{L}}^{\ast }\equiv \frac{\ell }{v_{h}(t)}\overline{%
\mathcal{L}},\hspace{0.25in}\zeta _{h}^{\ast }=\frac{\ell }{v_{h}(t)}\zeta
_{h}  \label{2.8}
\end{equation}%
If $\ell $ is the mean free path, then $s$ has the interpretation of the
average number of collisions in the interval $(0,t)$. The Liouville equation
for $\rho ^{\ast }$ then becomes%
\begin{equation}
\left( \partial _{s}+\overline{\mathcal{L}}^{\ast }\right) \rho ^{\ast }=0.
\label{2.9}
\end{equation}%
The definition of $\overline{\mathcal{L}}$ is such that its dimensionless
form, $\overline{\mathcal{L}}^{\ast }$, is independent of $s$. In this
representation of the Liouville equation, it is seen that the HCS is a
stationary solution.

It is possible to identify a set of important eigenvalues and eigenvectors
of the generator of dynamics $\overline{\mathcal{L}}^{\ast }$ by
differentiating (\ref{2.6}) with respect to its parameters $n,T,\mathbf{U}$.
Actually, it is convenient to change variables from $n_{h}$ and $T_{h}$ to $%
n_{h}$ and $\zeta _{h}\left( n_{h},T_{h}\right) $. The result is the set of
five eigenvalue equations%
\begin{equation}
\overline{\mathcal{L}}^{\ast }\Phi _{\alpha }=\lambda _{\alpha }^{\ast }\Phi
_{\alpha },  \label{2.10}
\end{equation}%
with the eigenvalues and eigenvectors given in terms of $\rho _{h}^{\ast }$
\begin{equation}
\lambda _{\alpha }^{\ast }=\left( 0,\frac{1}{2}\zeta _{h}^{\ast },-\frac{1}{2%
}\zeta _{h}^{\ast },-\frac{1}{2}\zeta _{h}^{\ast },-\frac{1}{2}\zeta
_{h}^{\ast }\right)  \label{2.11}
\end{equation}%
\begin{equation}
\Phi _{1}=\frac{\partial \rho _{h}}{\partial n_{h}}\mid _{\zeta _{h}},%
\hspace{0.25in}\Phi _{2}=\frac{\partial \rho _{h}}{\partial \zeta _{h}}\mid
_{n_{_{h}}},\hspace{0.25in}\Phi _{i=3,4,5}=\frac{\partial \rho _{h}}{%
\partial U_{i}}.  \label{2.12}
\end{equation}%
The significance of this lies in the fact that these eigenvalues are the
\textit{same} as those of the macroscopic balance equations for average
number density, energy density (or temperature), and momentum density (or
flow velocity) in the long wavelength limit. In this sense they represent
hydrodynamic modes at the level of the Liouville equation. This is not so
surprising since the corresponding modes in the elastic limit are those of
the corresponding conserved quantities, all with eigenvalues $0$. The
elastic limit results are recovered since in that limit $\zeta _{h}^{\ast
}\rightarrow 0$ and $\rho _{h}\rightarrow $ equilibrium distribution.

\section{Linear response}

\label{sec3}For molecular fluids Onsager's regression hypothesis states that
the dynamics of spontaneous fluctuations in an equilibrium fluid are the
same as those for the approach to equilibrium for weakly perturbed
non-equilibrium states. This has been formalized by the extensive
theoretical representations of linear response, and their exploration by
simulations and experiment \cite{Hansen,McL}. An extension of Onsager's
observation to granular fluids is described here, with special attention to
hydrodynamics for the large space and time scale non-equilibrium dynamics

\subsection{Response functions}

The objective is to describe the dynamics of the hydrodynamic fields due to
small initial perturbations of the HCS. These fields are chosen to be the
number density $n\left( \mathbf{r},t\right) $, cooling rate $\zeta \left(
\mathbf{r},t\right) $, and flow velocity $\mathbf{U}\left( \mathbf{r}%
,t\right) $
\begin{equation}
y_{\alpha }\left( \mathbf{r},t\right) \Leftrightarrow \left\{ n\left(
\mathbf{r},t\right) ,\zeta \left( \mathbf{r},t\right) ,U_{3}\left( \mathbf{r}%
,t\right) ,U_{4}\left( \mathbf{r},t\right) ,U_{5}\left( \mathbf{r},t\right)
\right\} .  \label{3.1}
\end{equation}%
The temperature has been replaced by the local cooling rate, $\zeta \left(
n,T\right) =c\left( n\right) T^{1/2}$, since the dimensionless macroscopic
balance equations are diagonal in this representation in the long wavelength
limit. The deviation of these fields from their values in the HCS are given
by%
\begin{equation}
\delta y_{\alpha }\left( \mathbf{r},t\right) =\int d\Gamma a_{\alpha }\left(
\Gamma ;\mathbf{r}\right) \left( \rho \left( \Gamma ,t\right) -\rho
_{h}\left( \Gamma ,t\right) \right) ,  \label{3.2}
\end{equation}%
where the phase functions $a_{\alpha }\left( \Gamma ;\mathbf{r}\right) $ are
linear combinations of those for the local density, energy density, and
momentum density; their dimensionless forms are given below. The initial
state is chosen to be a "local" HCS in the sense that the state is perturbed
only through the values of these hydrodynamic fields fields (see (\ref{2.4}%
))
\begin{equation}
\rho _{\ell h}\left( \left\{ y_{\alpha }\left( 0\right) \right\} \right)
\equiv \prod_{\beta =1}^{N}(lv(\mathbf{q}_{s},0))^{-3}\rho _{h}^{\ast
}\left( \{\frac{\mathbf{q}_{\alpha \gamma }}{l},\frac{\mathbf{v}_{\alpha }-%
\mathbf{U}\left( \mathbf{q}_{\alpha },0\right) }{v_{0}(\mathbf{q}_{\alpha
},0)}\},n\left( \mathbf{q}_{\alpha },0\right) l^{3}\right) .  \label{3.3}
\end{equation}%
This can be viewed as a state for which each small cell of the fluid is in a
HCS associate with the local values $\left\{ y_{\alpha }\left( \mathbf{r}%
,t\right) \right\} $. It is not a solution to the Liouville equation, but
rather a state whose average fields are specified as functions of $\mathbf{r}%
,t$. In this sense it is the analogue of the local equilibrium ensemble for
molecular fluids.

Use of this initial state in (\ref{3.2}) and expansion of $\rho _{\ell h}$
about the uniform values of $\left\{ y_{\alpha }\right\} $ for the HCS to
linear order gives the linear response%
\begin{equation}
\delta y_{\alpha }\left( \mathbf{r},t\right) =\int d\mathbf{r}^{\prime
}R_{\alpha \beta }\left( \mathbf{r,}t\mathbf{;r}^{\prime },0\right) \delta
y_{\beta }\left( \mathbf{r}^{\prime },0\right) .  \label{3.4}
\end{equation}%
The response functions are identified as%
\begin{equation}
R_{\alpha \beta }\left( \mathbf{r,r}^{\prime },t\right) =\int d\Gamma
a_{\alpha }(\Gamma ,\mathbf{r})e^{-\overline{L}t}\phi _{\beta }\left( \Gamma
;\mathbf{r}^{\prime }\right) .  \label{3.5}
\end{equation}%
with%
\begin{equation}
\phi _{\beta }\left( \Gamma ;\mathbf{r}\right) \equiv \sum_{i=1}^{N}\delta (%
\mathbf{q}_{i}-\mathbf{r})\frac{\partial \rho _{\ell h}\left( \left\{
y_{\alpha }\left( 0\right) \right\} \right) }{\partial y_{\alpha }\left(
\mathbf{q}_{i},0\right) }\mid _{\delta y=0}.  \label{3.6}
\end{equation}%
It is seen that the functions $\left\{ \phi _{\beta }\right\} $ are just the
densities associated with the hydrodynamic modes of (\ref{2.10}), $\int d%
\mathbf{r}\phi _{\beta }\left( \Gamma ;\mathbf{r}\right) =\Phi _{\alpha
}\left( \Gamma \right) .$ This is analogous to the linear response of a
molecular fluid to an initial local equilibrium state, where the
perturbations are proportional to the densities associated with the
conserved quantities (the long wavelength hydrodynamic modes).

It is appropriate at this point to introduce the dimensionless variables of
the last section. Also, since the reference HCS is uniform the response
function depends on $\mathbf{r,r}^{\prime }$ only through $\mathbf{r-r}%
^{\prime }$ and a Fourier representation is convenient%
\begin{equation}
\widetilde{f}(\mathbf{k}^{\ast })\equiv \int d\mathbf{r}^{\ast }e^{i\mathbf{k%
}^{\ast }\mathbf{\cdot r}^{\ast }}f(\mathbf{r}^{\ast }),\hspace{0.25in}f(%
\mathbf{r}^{\ast })=\frac{1}{V^{\ast }}\sum_{\mathbf{k}^{\ast }}e^{-i\mathbf{%
k}^{\ast }\mathbf{\cdot r}^{\ast }}f(\mathbf{k}^{\ast }).  \label{3.7a}
\end{equation}%
The dimensionless response functions are then%
\begin{equation}
\widetilde{R}_{\alpha \beta }^{\ast }\left( \mathbf{k}^{\ast };s\right)
\equiv \frac{1}{V^{\ast }}\int d\Gamma ^{\ast }\widetilde{a}_{\alpha }^{\ast
}(\Gamma ^{\ast },\mathbf{k}^{\ast })e^{-\overline{\mathcal{L}}^{\ast }s}%
\widetilde{\phi }_{\beta }^{\ast }\left( \Gamma ^{\ast };-\mathbf{k}^{\ast
}\right)  \label{3.8}
\end{equation}%
with%
\begin{equation}
\widetilde{a}_{\alpha }^{\ast }(\Gamma ^{\ast },\mathbf{k}^{\ast
})=\sum_{s=1}^{N}e^{i\mathbf{k}^{\ast }\mathbf{\cdot q}_{s}^{\ast
}}a_{\alpha }\left( \mathbf{v}_{s}^{\ast }\right) ,\hspace{0.25in}a_{\alpha
}\left( \mathbf{v}^{\ast }\right) \Leftrightarrow \left\{ 1,\frac{v^{\ast 2}%
}{3}-\frac{1}{2}+\frac{\partial \ln \zeta }{\partial n^{\ast }},\widehat{%
\mathbf{k}}^{\ast }\cdot \mathbf{v}^{\ast },\widehat{\mathbf{e}}_{1}\cdot
\mathbf{v}^{\ast },\widehat{\mathbf{e}}_{2}\cdot \mathbf{v}^{\ast }\right\} .
\label{3.9}
\end{equation}%
It follows from the definition of $\widetilde{R}_{\alpha \beta }^{\ast
}\left( \mathbf{k}^{\ast };s\right) $ at $s=0$ that $\left\{ \widetilde{a}%
_{\alpha }\right\} $ and $\left\{ \widetilde{b}_{\alpha }\right\} $ form a
bi-orthogonal set in the sense%
\begin{equation}
\frac{1}{V^{\ast }}\int d\Gamma ^{\ast }\widetilde{a}_{\alpha }^{\ast
}(\Gamma ^{\ast },\mathbf{k}^{\ast })\widetilde{\phi }_{\beta }^{\ast
}\left( \Gamma ^{\ast };-\mathbf{k}^{\ast }\right) =\delta _{\alpha \beta }.
\label{3.10}
\end{equation}

\subsection{Hydrodynamics}

If the fields obey a closed set of hydrodynamic equations on large space and
time scales, that dynamics must also be reflected in the dynamics of the
response functions (Onsager's regression relationship)%
\begin{equation}
\widetilde{R}^{\ast }\left( \mathbf{k}^{\ast };s\right) \rightarrow A(%
\mathbf{k}^{\ast })e^{-\lambda ^{\ast }\left( \mathbf{k}^{\ast }\right) s},
\label{3.11}
\end{equation}%
where $\lambda ^{\ast }\left( \mathbf{k}^{\ast }\right) $ is matrix whose
eigenvalues are those for the hydrodynamic modes. The matrix can be given a
precise representation in terms of the response functions
\begin{equation}
\lambda ^{\ast }\left( \mathbf{k}^{\ast }\right) =-\lim \left( \partial
_{s}R^{\ast }\right) R^{\ast -1}.  \label{3.28}
\end{equation}%
The limit indicated means long times and small $k^{\ast }$. A
straightforward expansion to order $k^{\ast 2}$ leads to the Green-Kubo
expressions for granular Navier-Stokes hydrodynamics. This will not be
repeated here, other than to note from (\ref{3.6}) that $\left\{ \widetilde{%
\phi }_{\beta }^{\ast }\left( \Gamma ^{\ast };-\mathbf{0}^{\ast }\right)
\right\} $ are the long wavelength hydrodynamic modes of the Liouville
equation so it follows directly that $\lambda _{\alpha \beta }^{\ast }\left(
\mathbf{0}\right) =\lambda _{\alpha }^{\ast }\delta _{\alpha \beta }$,%
\begin{equation}
\widetilde{R}_{\alpha \beta }^{\ast }\left( \mathbf{0};s\right) =e^{-\lambda
_{\alpha }^{\ast }s}\delta _{\alpha \beta },  \label{3.29}
\end{equation}%
with $\lambda _{\alpha }^{\ast }$ given by (\ref{2.11}).

\section{Mori identity and generalized hydrodynamic equations}

\label{sec4}Formally exact equations for the response functions can be
obtained by the projection operator method developed by Zwanzig and Mori
\cite{Zwanzig}. Its generalization to the granular response functions here
is straightforward. To simplify the notation a matrix representation will be
used and the asterisk and dependence on $\mathbf{k}$ suppressed for the
moment. Define the bi-linear functional $\left( x,\chi \right) $ by%
\begin{equation}
\left( x,\chi \right) =\frac{1}{V}\int d\Gamma x(\Gamma )\chi \left( \Gamma
\right) ,  \label{4.1}
\end{equation}%
where $x$ and $\chi $ are from the dual spaces of $a$ and $\phi $,
respectively. It is understood that dimensionless variables are used
throughout so the asterisk is also suppressed. The response function is then%
\begin{equation}
R(s)=\left( a\left( s\right) ,\phi \right) =\left( a,\phi \left( s\right)
\right) ,  \label{4.2}
\end{equation}%
with the phase space and Liouville dynamics defined by%
\begin{equation}
a\left( s\right) \equiv e^{\mathcal{L}s}a,\hspace{0.25in}\phi \left(
s\right) \equiv e^{-\overline{\mathcal{L}}s}\phi .  \label{4.3}
\end{equation}%
In this context $\mathcal{L}$ is the adjoint of $-\overline{\mathcal{L}}$.

Define the projection operators%
\begin{equation}
\mathcal{P}x=\left( x,\phi \right) a,\hspace{0.25in}\mathcal{P}^{\dagger
}\chi =\phi \left( a,\chi \right) .  \label{4.4}
\end{equation}%
From (\ref{3.10}), $\left( a,\phi \right) =1$ and it follows that $\mathcal{P%
}^{2}=\mathcal{P},$ $\mathcal{P}^{\dagger 2}=\mathcal{P}^{\dagger }$, and $%
\left( \mathcal{P}x,\chi \right) =\left( x,\mathcal{P}^{\dagger }\chi
\right) $. The response function has the representations%
\begin{equation}
R(s)=\left( \mathcal{P}a\left( s\right) ,\phi \right) =\left( a,\mathcal{P}%
^{\dagger }\phi \left( s\right) \right) .  \label{4.5}
\end{equation}%
Thus the response functions depend only on the projected dynamics. Mori's
identity is a decomposition of the time evolution of $a\left( s\right) $ or $%
\phi \left( s\right) $ into contributions from this projected dynamics and a
remainder orthogonal to it (see Appendix)%
\begin{equation}
a\left( s\right) =R(s)a+\int_{0}^{s}d\tau R(s-\tau )h\left( \tau \right) ,%
\hspace{0.25in}h\left( s\right) =e^{\mathcal{QLQ}s}\mathcal{QL}a,
\label{4.6}
\end{equation}%
\begin{equation}
\phi \left( s\right) =\phi R(s)+\int_{0}^{s}d\tau \gamma (s-\tau )R\left(
\tau \right) ,\hspace{0.25in}\gamma \left( s\right) =-\mathcal{Q}^{\dagger
}e^{-\mathcal{Q}^{\dagger }\overline{\mathcal{L}}\mathcal{Q}^{\dagger }s}%
\mathcal{Q}^{\dagger }\overline{\mathcal{L}}\phi ,  \label{4.7}
\end{equation}%
where $\mathcal{Q}=1-\mathcal{P}$. The equation for $R(s)$ now follows
directly from differentiating (\ref{4.2}) and use of Mori's identity%
\begin{equation}
\partial _{s}R(s)+\Omega R(s)+\int_{0}^{s}d\tau \Lambda \left( s-\tau
\right) R(\tau )=0,  \label{4.8}
\end{equation}%
The matrices generating the dynamics are%
\begin{equation}
\Omega =\left( a,\overline{\mathcal{L}}\phi \right) ,\hspace{0.25in}\Lambda
\left( s\right) =-\left( h(0),\gamma (s)\right) .  \label{4.9}
\end{equation}

Finally, differentiating the identities and use of (\ref{4.8}) leads to new
equations for $a(s)$ and $\phi (s)$
\begin{equation}
\partial _{s}a(s)+\Omega a(s)+\int_{0}^{s}d\tau \Lambda \left( s-\tau
\right) a(\tau )=h\left( s\right) ,  \label{4.10}
\end{equation}%
\begin{equation}
\partial _{s}\phi (s)+\phi (s)\Omega +\int_{0}^{s}d\tau \phi (\tau )\Lambda
\left( s-\tau \right) =\gamma \left( s\right) .  \label{4.11}
\end{equation}%
The left sides of the equations represent the macroscopic or average
dynamics of the response functions. The functions $h\left( s\right) $ and $%
\gamma \left( s\right) $ on the right sides are orthogonal to $a(s)$ and $%
\phi (s)$, respectively, and have the interpretation of sources for "noise".
Consequently, these equations have the interpretation of generalized
Langevin equations \cite{Zwanzig}. In this context, the second equality of (%
\ref{4.9}) represents a "fluctuation-dissipation" relation. All of the
results to this point are still exact.

\subsection{Hydrodynamics}

The relationship of the formally exact equation (\ref{4.8}) to hydrodynamics
is most easily discussed in terms of the Laplace transform of the response
functions%
\begin{equation}
\mathrm{R}(\mathbf{k},z)\equiv \int_{0}^{\infty }dse^{-zs}\widetilde{R}%
\left( \mathbf{k};s\right) ,\hspace{0.25in}\text{Re}\;z>z_{0},  \label{4.12}
\end{equation}%
where it is assumed that there exists some real $z_{0}$ such that $%
\widetilde{R}_{\alpha \beta }\left( \mathbf{k};s\right) \leq e^{-z_{0}s}$
for all $s$. Then (\ref{4.8}) gives%
\begin{equation}
\mathrm{R}(\mathbf{k},z)=\left( z+\Omega (\mathbf{k})+\overline{\Lambda }(%
\mathbf{k},z)\right) ^{-1},  \label{4.13}
\end{equation}%
where $\overline{\mathrm{\Lambda }}(\mathbf{k},z)$ is the Laplace transform
of $\mathrm{\Lambda }(\mathbf{k},s)$. The zeros of $\det \left( z+\Omega (%
\mathbf{k})+\overline{\mathrm{\Lambda }}(\mathbf{k},z)\right) $ define the
spectrum of $\mathrm{R}(\mathbf{k},z)$ and the complete dynamics of $%
\widetilde{R}\left( \mathbf{k};s\right) $. The spectrum for $\mathbf{k=0}$
is known from the exact result (\ref{3.29}) showing a set of poles at $%
z=-\lambda _{\alpha }.$ Assuming only continuity in $\mathbf{k}$ these poles
shift at finite $\mathbf{k}$ to the values $\lambda _{\alpha }\left( \mathbf{%
k}\right) $ determined from%
\begin{equation}
\det \left( -\lambda _{\alpha }\left( \mathbf{k}\right) +\Omega (\mathbf{k})+%
\overline{\mathrm{\Lambda }}(\mathbf{k},-\lambda _{\alpha }\left( \mathbf{k}%
\right) )\right) =0.  \label{4.14}
\end{equation}%
If the dependence on $\mathbf{k}$ is analytic then the solutions to (\ref%
{4.14}) can be constructed as a power series, $\lambda _{\alpha }\left(
\mathbf{k}\right) =\lambda _{\alpha }\mathbf{+}k\lambda _{\alpha
}^{(1)}+k^{2}\lambda _{\alpha }^{(2)}\mathbf{+..}$ The eigenvalues truncated
at order $k^{2}$ define the Navier-Stokes excitations.

The corresponding hydrodynamic equations are those of (\ref{3.11})
\begin{equation}
\left( \partial _{s}+\lambda \left( \mathbf{k}\right) \right) \widetilde{R}%
\left( \mathbf{k};s\right) =A\left( \mathbf{k}\right) ,  \label{4.16}
\end{equation}%
with the identification of $A\left( \mathbf{k}\right) $ as the residues of
the hydrodynamic poles and
\begin{equation}
\lambda \left( \mathbf{k}\right) =\Omega (\mathbf{k})+\overline{\mathrm{%
\Lambda }}(\mathbf{k},-\lambda \left( \mathbf{k}\right) )  \label{4.17}
\end{equation}%
The explicit Navier-Stokes form is obtained by expanding the expressions for
$\Omega (\mathbf{k})$ and $\overline{\Lambda }(\mathbf{k},-\lambda \left(
\mathbf{k}\right) )$ from (\ref{4.9}) to order $k^{2}$%
\begin{equation}
\lambda (\mathbf{k})\rightarrow \lambda +k\Omega ^{(1)}+k^{2}\left( \Omega
^{(2)}+\overline{\mathrm{\Lambda }}^{(2)}(z=-\lambda )\right) .  \label{4.18}
\end{equation}%
The terms of through order $k$ define the Euler hydrodynamics for a granular
fluid, while the contributions of order $k^{2}$ define the Navier-Stokes
order transport coefficients. The latter are the Green-Kubo expressions.

There are several important similarities and differences between the
hydrodynamics for granular and molecular fluids. At the Euler level, there
are the excitations $\lambda $ which lead to a long wavelength instability
not present for molecular fluids. The hydrostatic pressure has a dependence
on temperature and density determined by the reference HCS distribution
rather than the Gibbs distribution for a molecular fluid. Finally, there are
dissipative contributions from the dependence of the cooling rate on the
divergence of the flow field - a new transport coefficient.

At the Navier-Stokes level the viscous dissipation has the same form in both
cases. However, Fourier's law is modified by an additional dependence of the
heat flux on the gradient of the density. This can be traced to the failure
of Onsager's reciprocal relations, or time reversal invariance, in the
granular fluid. Finally, there are important differences in the form of the
Green-Kubo expressions for transport coefficients. As an explicit example,
the shear viscosity from (\ref{4.18}) is found to be \cite{Aparna}
\begin{equation}
\eta =-\frac{1}{V}\int d\Gamma \ T_{xy}\mathcal{M}_{xy}+\lim_{s\rightarrow
\infty }\int_{0}^{s}ds^{\prime }\int d\Gamma T_{xy}e^{-s\left( \overline{%
\mathcal{L}}+\frac{\zeta _{0}}{2}\right) }\Upsilon _{xy},  \label{4.19}
\end{equation}%
where the $T_{ij}$ is the usual volume integrated momentum flux tensor, and
the new flux $\Upsilon _{xy}$ and moment $\mathcal{M}_{xy}$
\begin{equation}
\Upsilon _{xy}=-\mathcal{Q}^{\dagger }\left( \overline{\mathcal{L}}+\frac{%
\zeta _{0}}{2}\right) \mathcal{M}_{xy},\hspace{0.25in}\mathcal{M}_{xy}=-%
\frac{1}{2}\sum_{r=1}^{N}\left( q_{r,x}\frac{\partial }{\partial
v_{r,y}^{\ast }}+q_{r,y}\frac{\partial }{\partial v_{r,x}^{\ast }}\right)
\rho _{h}.  \label{4.20}
\end{equation}%
There are two parts, a time integral of a flux-flux correlation function as
in the usual expressions for molecular fluid, and a time-independent
contribution. Both terms occur in the elastic limit, with the latter being a
peculiarity of the hard sphere interaction. However, for the granular fluid
this expression occurs even for continuous particle interactions and is due
to the absence of time reversal invariance. Also in the elastic limit the
flux $\Upsilon _{xy}\rightarrow $ $T_{xy}$ and the correct Green-Kubo
expression for the viscosity of a hard sphere fluid is regained. However,
the granular fluid flux $\Upsilon _{xy}\neq T_{ij}$. Instead one of the
fluxes is associated with the local conserved densities $\widetilde{a}%
_{\alpha }$ of (\ref{3.9}) while the other flux is associated with the
bi-orthogonal densities $\widetilde{\phi }_{\beta }$ of (\ref{3.6}).
Finally, it is noted that the flux-flux correlation function has the
invariant subspace of the dynamics projected out, a necessary condition for
the convergence of the time integral for large $s$.

\section{Kinetic theory of response functions}

\label{sec5}Practical evaluation of the response functions for all $\mathbf{k%
},s$ is a formidable task for both molecular and granular fluids. In the
former case, phenomenological approaches based on modeling "memory
functions" in terms of a few moments has met with considerable success \cite%
{Boon}. More controlled and detailed descriptions are provided by kinetic
theory. In this section the representation of response functions appropriate
for such methods is given \cite{Aparna2}. Subsequently, two practical
approximations are described.

The Fourier transformed response function can be written in the alternative
form%
\begin{equation}
\widetilde{R}_{\alpha \beta }\left( \mathbf{k};s\right) =\int d\Gamma
\widetilde{a}_{\alpha }(\Gamma ,\mathbf{k})e^{-\overline{\mathcal{L}}s}\phi
_{\beta }\left( \Gamma ;\mathbf{r=0}\right) .  \label{5.1}
\end{equation}%
Then, since the local densities $\widetilde{a}_{\alpha }(\Gamma ,\mathbf{k})$
are sums of single particle functions (see (\ref{3.9})), integration over
all degrees of freedom except those for one particle can be performed to get
a representation in the single particle phase space%
\begin{equation}
\widetilde{R}_{\alpha \beta }\left( \mathbf{k};s\right) =n\int d\mathbf{v}%
_{1}d\mathbf{q}_{1}e^{i\mathbf{k\cdot q}_{1}}a_{\alpha }\left( \mathbf{v}%
_{1}\right) \phi _{\alpha }^{(1)}\left( x_{1};\mathbf{r=0},s\right) ,
\label{5.2}
\end{equation}%
The reduced density $\phi _{\alpha }^{(1)}\left( x_{1};\mathbf{0},s\right) $
is a member of a set defined by%
\begin{equation}
n^{m}\phi _{\alpha }^{(m)}\left( x_{1},\cdots ,x_{m};\mathbf{0},s\right) =%
\frac{N!}{\left( N-m\right) !}\int dx_{m+1}..dx_{N}e^{-\overline{\mathcal{L}}%
s}\phi _{\alpha }\left( \Gamma ;\mathbf{0}\right) .  \label{5.4}
\end{equation}%
Here $x_{i}=\mathbf{q}_{i},\mathbf{v}_{i}$ denotes a point in the six
dimensional phase space for a single particle. These functions obey a
hierarchy of equations obtained by direct differentiation. The first of
these equations describes the dynamics of the single particle function $\phi
_{\alpha }^{(1)}\left( x_{1};\mathbf{0},s\right) $
\begin{equation}
\left( \partial _{s}+\frac{1}{2}\zeta \left( 3+\mathbf{v}_{1}\cdot \nabla _{%
\mathbf{v}_{1}}\right) +\mathbf{v}_{1}\cdot \nabla _{\mathbf{q}_{1}}\right)
\phi _{\alpha }^{(1)}=n\sigma ^{2}\int dx_{2}\,\overline{T}(1,2)\phi
_{\alpha }^{(2)}(x_{1},x_{2},s),  \label{5.5}
\end{equation}%
where $\overline{T}(1,2)$ is the binary collision operator describing
velocity changes due to hard sphere interactions (its detailed form \cite%
{Kandrup,vanN01} is not needed here). Equation (\ref{5.5}) shows the
coupling of $\phi _{\alpha }^{(1)}$ to the two particle function $\phi
_{\alpha }^{(2)}$, which in turn obeys an equation coupled to $\phi _{\alpha
}^{(2)};$ hence the structure of a hierarchy. A closed \emph{kinetic equation%
} for $\phi _{\alpha }^{(1)}$ results from this first hierarchy equation if $%
\phi _{\alpha }^{(2)}$ can be expressed approximately as a functional of $%
\phi _{\alpha }^{(1)}$
\begin{equation}
\phi _{\alpha }^{(2)}(x_{1},x_{2},s)\simeq \Phi (x_{1},x_{2},s\mid \phi
_{\alpha }^{(1)}).  \label{5.6}
\end{equation}%
Use of this functional in (\ref{5.4}) gives the desired kinetic equation,
and its solution gives the response function via (\ref{5.2}).

Generally, such a functional relationship entails limitations of space and
time scales, density or other conditions for validity. An example is the
Markovian approximation. This represents the functional by its exact form at
$s=0$
\begin{equation}
\Phi (x_{1},x_{2},s\mid \cdot )\rightarrow \Phi (x_{1},x_{2},s=0^{+}\mid
\cdot ).  \label{5.7}
\end{equation}%
This is a type of mean field approximation, representing the collisional \
process by its average value in the initial state of the system. It is
therefore exact at short times but only approximate thereafter. For the hard
sphere interactions considered here, it describes the non-trivial binary
collisions, modified by correlations with the environment. It neglects
dynamical correlations that develop over longer times. The detailed form of
the Markovian approximation is worked out elsewhere \cite{Aparna2} and only
the result is quoted here. Equation (\ref{5.5}) becomes the kinetic equation
\begin{equation}
\left( \partial _{s}+\frac{1}{2}\zeta \nabla _{\mathbf{v}_{1}}\cdot \mathbf{v%
}_{1}+\mathbf{v}_{1}\cdot \nabla _{\mathbf{q}_{1}}+\widehat{I}\right) \phi
_{\alpha }^{(1)}=0,  \label{5.8}
\end{equation}%
where the collision operator is given by%
\begin{eqnarray}
\widehat{I}\phi _{\alpha }^{(1)} &=&\int dx_{s+1}\,\overline{T}(1,2)\Phi
(x_{1},x_{2},s=0^{+}\mid \phi _{\alpha }^{(1)})  \notag \\
&=&-\int dx_{2}\,\overline{T}(x_{1},x_{2})g_{h}^{(2)}(x_{1},x_{2})\left[
f_{h}^{(1)}(v_{1})\phi _{\alpha }^{(1)}(x_{2},s)+f_{h}^{(1)}(v_{2})\phi
_{\alpha }^{(1)}(x_{1},s)\right]  \notag \\
&&-\int d{\mathbf{q}}_{2}c_{\lambda }({\mathbf{v}}_{1},{\mathbf{q}}_{12})%
\frac{1}{n_{h}}\int d{\mathbf{v}}_{2}a_{\lambda }\left( {\mathbf{v}}%
_{2}\right) \phi _{\alpha }^{(1)}(x_{2},s).  \label{5.9}
\end{eqnarray}%
The function $g_{h}^{(2)}(x_{1},x_{2})$ describes the pair correlations in
the HCS (including velocity correlations), and $c_{\lambda }({\mathbf{v}}%
_{1},{\mathbf{q}}_{12})$ is related to its functional derivative with
respect to the local hydrodynamic fields $y_{\lambda }({\mathbf{q}}_{2})$
\begin{equation}
c_{\lambda }({\mathbf{v}}_{1},{\mathbf{q}}_{12})=\int dx\overline{T}%
(x_{1},x)f_{h}^{(1)}(v_{1})f_{h}^{(1)}(v)\left[ \frac{\delta g_{\ell
h}^{(2)}[x_{1},x|\delta y]}{\delta y_{\lambda }({\mathbf{q}}_{2})}\right]
_{\delta y=0}.  \label{5.10}
\end{equation}%
The corresponding response functions are determined from the Fourier
transform of this equation, with the result%
\begin{equation}
\widetilde{R}_{\alpha \beta }\left( \mathbf{k};s\right) =n\int d\mathbf{v}%
_{1}a_{\alpha }\left( \mathbf{v}_{1}\right) e^{\left( i\mathbf{k\cdot v}-%
\widetilde{I}({\mathbf{k}})-\frac{1}{2}\zeta \nabla _{\mathbf{v}_{1}}\cdot
\mathbf{v}_{1}\right) s}\frac{\delta f_{\ell h}^{(1)}}{\delta y_{\beta }}%
\mid _{\delta y=0},  \label{5.11}
\end{equation}%
where $\widetilde{I}({\mathbf{k}})$ is the Fourier representation for the
collision operator $\widehat{I}$, and the initial condition is the
functional derivative of the single particle distribution $f_{\ell h}^{(1)}$
determined from the local HCS distribution (\ref{3.3}).

The Markovian approximation leads to a practical result, without \emph{a
priori} limitations on the density, or length scale. Thus it can be applied
to both dilute and dense fluid conditions, from hydrodynamic length scales
to those smaller than the grain size. In the elastic limit it reduces to the
generalized Enskog approximation that has met with great success in that
diverse context \cite{Hansen,Boon}. For Granular fluids, the required input
property $g_{h}^{(2)}(x_{1},x_{2})$ is not known and hence most applications
to date have used the further approximation of neglecting the velocity
correlations. Still, the results provide an accurate and practical extension
of the Boltzmann results to finite density gases.

\section{Response at Low Density}

A second example of a kinetic theory description for response can be
obtained more systematically using a dimensionless density as expansion
parameter. The starting point is the hierarchy for the reduced distribution
functions $f^{(m)}(x_{1},\cdots ,x_{m},s)$ associated $N$ particle
distribution function evolving from the initial local HCS distribution (\ref%
{3.3})%
\begin{equation}
n^{m}f^{(m)}\left( x_{1},\cdots ,x_{m};s\right) =\frac{N!}{\left( N-m\right)
!}\int dx_{m+1}..dx_{N}e^{-\overline{\mathcal{L}}s}\rho _{\ell h}\left(
\left\{ y_{\alpha }\left( 0\right) \right\} \right) .  \label{6.1}
\end{equation}%
It follows that the function $\phi _{\alpha }^{(1)}\left( x_{1};\mathbf{r=0}%
,s\right) $ that determines the response functions in (\ref{5.2}) is
obtained from $f^{(1)}(x_{1},s)$ by functional differentiation
\begin{equation}
\phi _{\alpha }^{(1)}\left( x_{1};\mathbf{r=0},s\right) =\left[ \frac{\delta
f^{(1)}(x_{1};s)}{\delta y_{\alpha }(\mathbf{0})}\right] _{\delta y=0}.
\label{6.2}
\end{equation}%
The analysis proceeds by first finding a kinetic equation for $f^{(1)}$ and
then determining the corresponding kinetic equation for $\phi _{\alpha
}^{(1)}$ by functional differentiation. Only an outline of the results is
provided here, with further details \cite{Dufty09} to be given elsewhere.

The first hierarchy equation is \cite{hierarchy}%
\begin{equation}
\left( \partial _{s}+\frac{1}{2}\zeta \nabla _{\mathbf{v}_{1}}\cdot \mathbf{v%
}_{1}+\mathbf{v}_{1}\cdot \nabla _{\mathbf{q}_{1}}\right)
f^{(1)}(x_{1};s)=\int dx_{2}\,\overline{T}(1,2)f^{(2)}(x_{1},x_{2};s).
\label{6.3}
\end{equation}%
A kinetic equation is obtained by finding a functional relationship $%
f^{(2)}(x_{i},x_{j};s)=F^{(2)}(x_{i},x_{j},s\mid f^{(1)})$. This is
accomplished as an expansion in the parameter $\sigma /\ell $, where $\sigma
$ is the grain diameter and $\ell =1/n\sigma ^{2}$ is the mean free path,
for $\sigma /\ell <<1$. The hierarchy equations can then be solved formally
for $f^{(m)}$ as a power series in this small parameter. The solutions can
be inverted to give a representation of $f^{(m)}$ in terms of $f^{(1)}$ and
the initial conditions. The result is

\begin{equation}
F^{(2)}(x_{1},x_{2};s\mid f^{(1)})=f^{(1)}(x_{1};s)f^{(1)}(x_{2};s)+\left(
\frac{\sigma }{\ell }\right) ^{2}C^{(2)}(x_{1},x_{2};s\mid f^{(1)})+..
\label{6.4}
\end{equation}%
where $C^{(2)}$ is determined from
\begin{equation}
\left( \partial _{s}+\mathcal{L}\left( 1\mid f^{(1)}\right) +\mathcal{L}%
\left( 2\mid f^{(1)}\right) \right) C^{(2)}(x_{1},x_{2};s\mid f^{(1)})=%
\overline{T}(1,2)f^{(1)}(x_{1};s)f^{(1)}(x_{2};s).  \label{6.5}
\end{equation}%
The single particle generator $\mathcal{L}\left( 1\mid f^{(1)}\right) $ is
\begin{eqnarray}
\mathcal{L}\left( 1\mid f^{(1)}\right) X(1) &=&\left( \mathbf{v}_{1}\cdot
\nabla _{\mathbf{q}_{1}}+\frac{1}{2}\zeta \nabla _{\mathbf{v}_{1}}\cdot
\mathbf{v}_{1}\right) X(1)  \notag \\
&&-\int dx_{3}\,\overline{T}(1,3)\left(
f^{(1)}(x_{1};s)X(x_{3})+X(x_{1})f^{(1)}(x_{3};s)\right) .  \label{6.6}
\end{eqnarray}%
These results determine the functional $F^{(2)}(x_{1},x_{2};s\mid \cdot )$
exactly to order $\left( \sigma /\ell \right) ^{2}$. For the special case $%
\delta y_{\alpha }=0$ in (\ref{6.1}), these equations become time
independent and determine the one and two particle reduced distribution
functions for the HCS at low density. Closely related equations have been
studied to determine pair correlations in the HCS \cite{Brey04}

To expose the content of this result, the linear equation for $C^{(2)}$ can
be solved
\begin{eqnarray}
C^{(2)}(x_{1},x_{2};s &\mid &f^{(1)})=\mathcal{U}(s,0)\left[ f_{\ell
h}^{(2)}(x_{1},x_{2})-f_{\ell h}^{(1)}(x_{1})f_{\ell h}^{(1)}(x_{1})\right]
\notag \\
&&+\int_{0}^{s}ds^{\prime }\mathcal{U}(s,s^{\prime })\overline{T}%
(1,2)f^{(1)}(x_{1};s^{\prime })f^{(1)}(x_{2};s^{\prime }),  \label{6.7}
\end{eqnarray}%
where the solution operator $\mathcal{U}(s,s^{\prime })$ obeys the equation%
\begin{equation}
\left( \partial _{s}+\mathcal{L}\left( x_{1}\mid f^{(1)}\right) +\mathcal{L}%
\left( x_{2}\mid f^{(1)}\right) \right) \mathcal{U}(s,s^{\prime })=0,\hspace{%
0.3in}\mathcal{U}(s^{\prime },s^{\prime })=1.  \label{6.8}
\end{equation}%
With these results, the first hierarchy equation (\ref{6.3}) becomes
\begin{equation*}
\left( \partial _{s}+\frac{1}{2}\zeta \nabla _{\mathbf{v}_{1}}\cdot \mathbf{v%
}_{1}+\mathbf{v}_{1}\cdot \nabla _{\mathbf{q}_{1}}\right)
f^{(1)}(x_{1};s)=\int dx_{2}\,\overline{T}%
(1,2)f^{(1)}(x_{1};s)f^{(1)}(x_{2};s)
\end{equation*}%
\begin{eqnarray}
&&+\left( \frac{\sigma }{\ell }\right) ^{2}\int_{0}^{s}ds^{\prime }\int
dx_{2}\,\overline{T}(1,2)\mathcal{U}(s,s^{\prime })\overline{T}%
(1,2)f^{(1)}(x_{1};s^{\prime })f^{(1)}(x_{2};s^{\prime })  \notag \\
&&+\left( \frac{\sigma }{\ell }\right) ^{2}\int dx_{2}\,\overline{T}(1,2)%
\mathcal{U}(s,0)\left[ f_{\ell h}^{(2)}(x_{1},x_{2})-f_{\ell
h}^{(1)}(x_{1})f_{\ell h}^{(1)}(x_{1})\right]  \label{6.9}
\end{eqnarray}%
To lowest order, the first term on the right side gives the granular
nonlinear Boltzmann kinetic equation. This approximation has been studied in
some detail for low density granular gases \cite{Poschel,Poschel1}. The
second term on the right side describes "ring collisions", which are
dynamically correlated recollisions among pairs due to sequential collisions
with many other particles. Finally, the last term describes related
dynamical correlations due to initial static correlations in the initial
local HCS.

The \emph{linear} kinetic equation for $\phi _{\alpha }^{(1)}$ to determine
the response functions follows from (\ref{6.2}), functional differentiation
of (\ref{6.9}), and setting $\delta y_{\alpha }=0$. It has the structural
form%
\begin{equation}
\left( \partial _{s}+\mathcal{L}_{h}\left( x_{1}\right) \right) \phi
_{\alpha }^{(1)}\left( x_{1};s\right) +\int_{0}^{s}ds^{\prime }\mathcal{R}%
\left( s-s^{\prime }\right) \phi _{\alpha }^{(1)}\left( x_{1};s^{\prime
}\right) =I_{\alpha }\left( x_{1};s\right) ,  \label{6.10}
\end{equation}%
where the source $I_{\alpha }$ is due to the evolution of initial pair
correlations, inherited from the last term of (\ref{6.2}). The operator $%
\mathcal{L}_{h}\left( 1\right) $ is the generator for dynamics of the linear
granular Boltzmann equation

\begin{eqnarray}
\mathcal{L}_{h}\left( x_{1}\right) X(x_{1}) &=&\left( \mathbf{v}_{1}\cdot
\nabla _{\mathbf{q}_{1}}+\frac{1}{2}\zeta _{h}\left( 3+\mathbf{v}_{1}\cdot
\nabla _{\mathbf{v}_{1}}\right) \right) X(x_{1})  \notag \\
&&-\int dx_{3}\,\overline{T}(1,3)\left(
f_{h}^{(1)}(v_{1})X(x_{3})+X(x_{1})f_{h}^{(1)}(v_{3})\right) .  \label{6.11}
\end{eqnarray}%
It can be shown that $\mathcal{L}_{h}\left( 1\right) $ is the low density
limit of the generator for the Markovian approximation (\ref{5.8}). The last
term on the left side of (\ref{6.10}) represents the correlated collisions
generated by the linear ring operator $\mathcal{R}\left( s\right) $, and the
source on the right side is due to the dynamical evolution of initial pair
correlations. Both of these terms are proportional to $\left( \sigma /\ell
\right) ^{2}$.Their detailed forms are lengthy and will not be given here.
In the elastic limit, they reduce to the ring kinetic theory for response
functions of a molecular fluid.

The most interesting feature of the ring operator is its description of a
mechanism for slow dynamics not present at the level of the Boltzmann
equation. It is due to a coupling of hydrodynamic modes that originates from
the generator $e^{-\left( \mathcal{L}_{h}\left( x_{1}\right) +\mathcal{L}%
_{h}\left( x_{2}\right) \right) s}$ in the solution to (\ref{6.8}). It is
well known that the hydrodynamic modes of a molecular fluid are contained in
the spectrum of the linearized Boltzmann operator, and a similar result has
been proved for the linearized granular Boltzmann equation \cite{Brey05}.
Thus in general the action of this operator generates the exponential decay
of two hydrodynamic modes at all possible wavelengths, leading to slow decay
(algebraic, logarithmic) of response functions for molecular fluids. Similar
slow decay has been predicted on more phenomenological "mode coupling"
methods for the related flux correlation functions in granular fluids \cite%
{Hayakawa07}. The above kinetic equation provides the means to investigate
this mechanism in more detail and broader context.

A Laplace transform of (\ref{6.10}) and its formal solution leads to the
Fourier and Laplace transformed response functions in the form%
\begin{equation}
\mathrm{R}(\mathbf{k},z)=n\int d\mathbf{v}a(\mathbf{v})\left( z+\mathcal{L}%
_{h}\left( \mathbf{k}\right) +\overline{\mathcal{R}}\left( \mathbf{k,}%
z\right) \right) ^{-1}\left( \Phi ^{(1)}(\mathbf{v})+\overline{I}_{0}\left(
\mathbf{v},\mathbf{k},z\right) \right) .  \label{6.12}
\end{equation}%
where $\Phi ^{(1)}(\mathbf{v})$ is the reduced single particle function
determined from the global invariant $\Phi $ in (\ref{2.12}). An important
consequence of the eigenvalue equation (\ref{2.10}) for the Liouville
equation is that these same long wavelength eigenvalues are poles of the
resolvent operator in (\ref{6.12})%
\begin{equation}
\left( \mathcal{L}_{h}\left( \mathbf{0}\right) +\overline{\mathcal{R}}\left(
\mathbf{0,}-\lambda _{\alpha }\right) \right) \Phi _{\alpha }^{(1)}=\lambda
_{\alpha }\Phi _{\alpha }^{(1)}.  \label{6.13}
\end{equation}%
Assuming analyticity in the wavevector $\mathbf{k}$ this assures the
existence of hydrodynamic excitations in the spectrum of $\mathcal{L}%
_{h}\left( \mathbf{k}\right) +\overline{\mathcal{R}}\left( \mathbf{k,}%
z\right) $. Then by perturbation theory \cite{Brey05} the hydrodynamic
modes, including effects of the ring collisions, can be explored through
Navier-Stokes order.

\section{Summary}

The presentation here has given an overview of linear response functions for
excitations in the HCS or a granular fluid. The HCS is the homogeneous state
of an isolated fluid, corresponding to the equilibrium state or a molecular
fluid, and is perhaps the simplest case to consider. The response for states
of more practical interest (steady shear or other driven steady states) is
considerably more complex, and relatively little is known even in the case
of molecular fluids. Here, an attempt was made to show the structural
features that are the same for granular and molecular fluids, while
attending to their differences in detail.

Although the utility of these functions to describe hydrodynamic response
has been emphasized, there application for shorter length and time scale
phenomena should be recognized. There are many methods for practical
evaluation suggested by experience with molecular fluids. The kinetic theory
approach has been illustrated here with the Markovian approximation that
encompasses all length and time scales and a range of densities well beyond
the Boltzmann limit. In addition, systemmatic corrections to the Boltmann
limit have been described for low density granular gases to include the
novel effects of correlated many-body collisions (rings). Still to be
explored are the consequences of this low density kinetic theory,
particularly the consequences of hydrodynamic mode coupling in the spectrum
of the ring operator. While this is well known for molecular fluids, there
is a qualitative difference in the hydrodynamic modes for granular gases.
Some of the modes are unstable (in dimensionless form) and there is the
potential for a qualitative difference in the hydrodynamic dispersion
relations determined from $\mathrm{R}(\mathbf{k},z)$. This occurs for
molecular fluids in two dimensions, due to mode coupling effects, and may be
more significant for granular fluids in all dimensions due to the unstable
modes. The detailed evaluation of (\ref{6.12}) will clarify this possibility.

\section{Acknowledgements}

The author is indebted to collaborators J. J. Brey, A. Baskaran, and J.
Lutsko with whom much of this research has been performed. The partial
support of the Yukawa International Program for Quark-Hadron Sciences
(YIPQS), and the Department of Energy award DE-FG02-07ER54946 is gratefully
acknowledged.

\appendix

\section{Mori's identity}

The projection operators of (\ref{4.4}) are $\mathcal{P}x=\left( x,\psi
\right) a,\hspace{0.25in}\mathcal{P}^{\dagger }\chi =\phi \left( a,\chi
\right) .$ Let $\mathcal{Q}=1-\mathcal{P}$ and $\mathcal{Q}^{\dagger }=1-%
\mathcal{P}^{\dagger }$ denote the corresponding orthogonal projections.
Then $a\left( s\right) $ can be decomposed as
\begin{equation}
a\left( s\right) =\left( \mathcal{P}+\mathcal{Q}\right) a\left( s\right)
=R(s)a+a_{\perp }(s),  \label{a.2}
\end{equation}%
where $a_{\perp }(s)=\mathcal{Q}a(s)\mathcal{Q}$. Next operator with $%
\mathcal{Q}$ on the equation of motion $\left( \partial _{s}-\mathcal{L}%
\right) a\left( s\right) =0$, to get
\begin{equation}
\left( \partial _{s}-\mathcal{QLQ}\right) a_{\perp }(t)=\mathcal{QLP}%
a(t)=R(s)\mathcal{QL}a.  \label{a.3}
\end{equation}%
The solution is%
\begin{equation}
a_{\perp }(t)=\int_{0}^{s}d\tau R(s-\tau )e^{\widehat{\mathcal{L}}\tau }%
\mathcal{QL}a,  \label{a.4}
\end{equation}%
and its substitution into (\ref{a.2}) gives Mori's identity (\ref{4.6}). A
similar analysis for $\psi \left( s\right) $ gives the identity (\ref{4.7}).

Finally, differentiating the response function and using Mori's identity
gives
\begin{eqnarray}
\partial _{s}R(s) &=&R(s)\left( \mathcal{L}a,\phi \right) +\int_{0}^{s}d\tau
R(s-\tau )\left( \mathcal{L}h\left( \tau \right) ,\phi \right)  \notag \\
&=&-\left( a,\overline{\mathcal{L}}\phi \right) R(s)-\int_{0}^{s}d\tau
\left( a,\overline{\mathcal{L}}\gamma (s-\tau )\right) R\left( \tau \right)
\label{a.5}
\end{eqnarray}%
The first equality comes from using the form $\left( a\left( s\right) ,\phi
\right) $ while the second results from the form $\left( a,\phi \left(
s\right) \right) $.

\end{document}